\documentclass[sn-mathphys,Numbered,scrartcl]{sn-jnl}

\usepackage{enumitem}
\usepackage{graphicx}%
\usepackage{multirow}%
\usepackage{amsmath,amssymb,amsfonts}%
\usepackage{amsthm}%
\usepackage{mathrsfs}%
\usepackage[title]{appendix}%
\usepackage{xcolor}%
\usepackage{textcomp}%
\usepackage{manyfoot}%
\usepackage{booktabs}%
\usepackage{algorithm}%
\usepackage{algorithmicx}%
\usepackage{algpseudocode}%
\usepackage{listings}%
\usepackage{pythontex}
\usepackage{listings}
\usepackage{multicol}
\usepackage{natbib}
\usepackage{url}
\usepackage{hyperref}

\definecolor{dkgreen}{rgb}{0,0.6,0}
\definecolor{gray}{rgb}{0.5,0.5,0.5}
\definecolor{mauve}{rgb}{0.58,0,0.82}

\begin{document}

\title[Article Title]{Playing Lato-lato is Difficult and This is Why}


\author[1]{\fnm{Fansen Candra} \sur{Funata}}\email{fansencandra.funata@gmail.com}
\author*[2]{\fnm{Zainul} \sur{Abidin}}\email{zainul.abidin@yayasansimetri.or.id}


\affil[1]{\orgdiv{XI Natural Science}, \orgname{Darma Yudha Senior High School}, \orgaddress{\street{189 S.M. Amin Street}, \city{Pekanbaru}, \postcode{28292}, \state{Riau}, \country{Indonesia}}}

\affil*[2]{Simetri Foundation, Tangerang, Indonesia 15334}


\label{Abstract}\abstract{Lato-lato, a pendulum-based toy gaining popularity in Indonesian playgrounds, has sparked interest with competitions centered around maintaining its oscillatory motion. While some find it easy to play, the challenge lies in sustaining the oscillation, particularly in maintaining both "up and down collisions." Through a Newtonian dynamics numerical analysis using Python (code by ChatGPT), this study identifies two equilibrium phases - phase 1, characterized by normal pendulum motion, and phase 2, the double collision mode - by using the driven oscillation model. In addition, further analysis and discussion are done using the obtained numeric data. The difficulty in remaining in phase 2 highlights the intricate hand-eye coordination required, shedding light on the toy's appeal and the skill it demands.}




\maketitle

\section{Introduction}\label{sec1}

Lato-lato, or better known as "clacker balls" has existed since the 1960s, originally made of tempered glass. Due to safety issues, the balls are then changed to be made out of plastic \cite{Clacker}. 
It recently regained its popularity, especially in Indonesia, because it was played by the president of Indonesia \cite{CNNIndonesia.2022}. Its appearance in social media, such as TikTok made it even more viral, with videos soaring to millions of views. Furthermore, playing this game positively impacts children's behaviour, increasing the frequency of their social interaction\cite{udasmoro2022impact}. \newline
Previously some reports and articles have been written revolving around lato-lato. However, most of them seem to focus on how educators can implement lato-lato in teaching physics\cite{akhsan2023newtonian}\cite{Wibowo_2023}\cite{Wibowo_2024}. As we all know, the physical essence of the toy \textit{lato-lato} lies in the law of momentum conservation; the collisions that occur between the plastic spheres are such that \(p_i=p_f\)\cite{article}\cite{Cross_2021}. However, the dynamics of the lato-lato itself (\ref{subsectionpd}), involve gruesome mathematics, as shown by Bartucelli \textit{et al}\cite{bartucelli2}.  In this paper, we will use the concept of \textbf{phase diagrams} to explain why the lato-lato is such a difficult game. Moreover, A diagram of the \textit{Amplitude Conditions}\footnote{Variable defined to show which phase the pendulum is in} will be shown against the initial boundary conditions, \(\theta_0\) and \(\dot\theta_0\). Further, we model the lato-lato as a 2 stick pendulum joint at both of its free ends. 
\begin{figure}[H]
    \centering
    \includegraphics[width=0.86\textwidth,clip=]{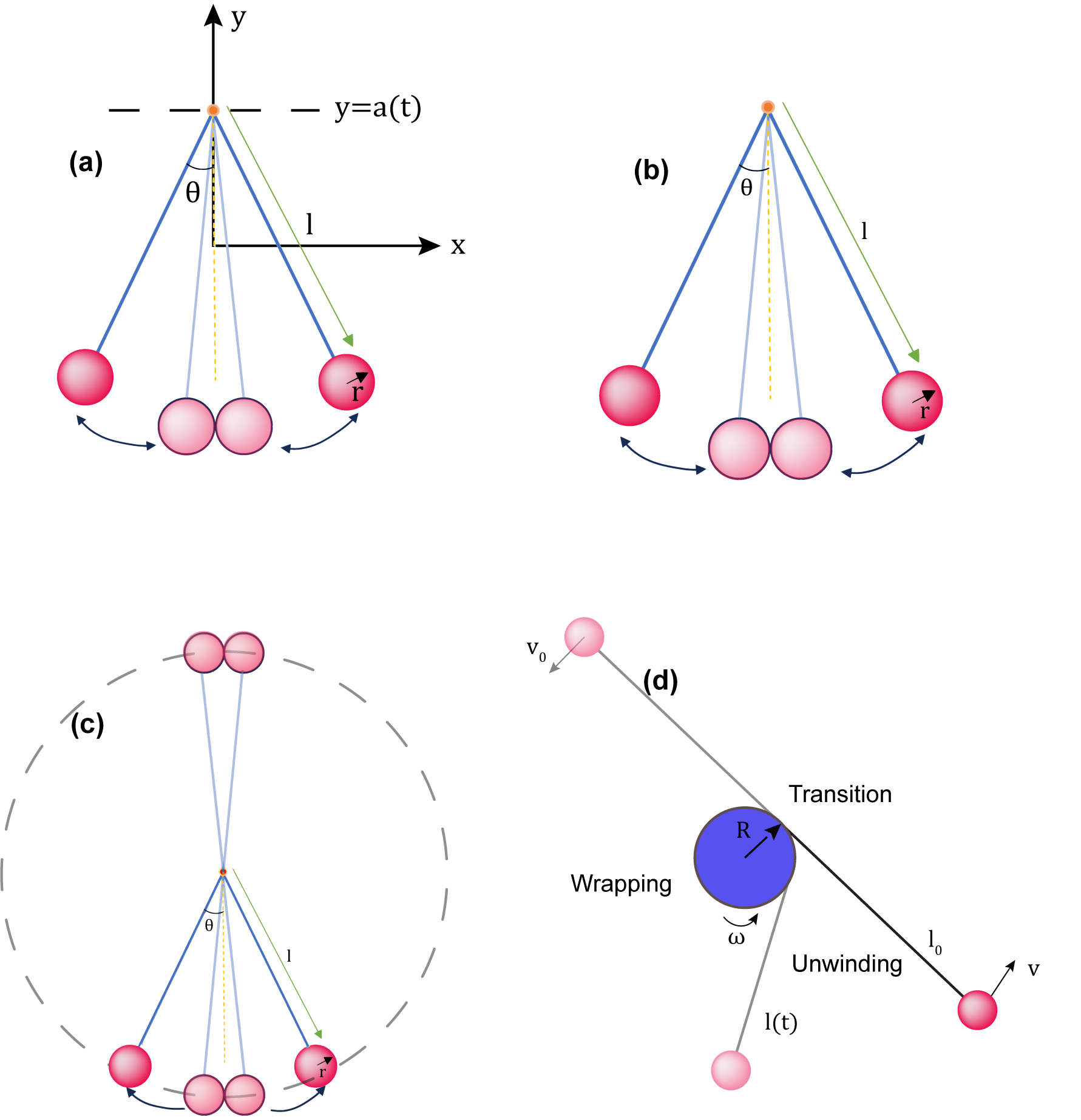}
    \caption{\textbf{(a)} shows the model of the lato-lato used throughout our analysis. We use the approximation \(r<<l\), hence the sphere may be considered a point mass. \textbf{(b)} Represents the motion of "phase 1", and \textbf{(c)} represents the motion of "phase 2". Figure \textbf{(d)} shows a special play style called the "tornado style" (more about it in \hyperref[subsec2.6]{Subsection 2.6})}
    \label{fig:enter-label1}
\end{figure}

\section{Formula Derivations}\label{sec2}
The concept involved in playing with this toy lies mainly in Newton's laws and conservation of momentum and energy\cite{morin2008introduction}. Some of the subsections we provide here serve as preliminary materials to aid the readers in grasping the materials as a whole (\hyperref[subsection1]{Section 2.1} and \hyperref[subsection2.2]{Section 2.2}).\newline
The formulas used are as follows
\subsection{Single Pendulum Equation of Motion}\label{subsection1}
We begin with a single pendulum case with its energy given as
\begin{equation}
    \label{energy 1}
    E=mgl(1-\cos\theta)+ml^2\frac{\dot{\theta}^2}{2}
\end{equation}
Where \(E\) represents the energy of the system, \(m\) represents the mass of the bob, \(g\) represents the gravitational acceleration (\(9.8\) \(m/s^2\)), \(l\) represents the length of the pendulum, \(\theta\) represents the deviation angle with respect to the y axis and \(\dot\theta\) represents the derivative with respect to time, as shown in \hyperref[fig:enter-label1]{Figure 1}.\newline\newline
Without energy loss, we consider three separate cases, which are \textbf{(a)} \(E<<2mgl\), \textbf{(b)} \(E=2mgl\) and \textbf{(c)} \(E>2mgl\). Here, we ought to find the angular momentum \(L\) of the system to obtain plots of the system's phase diagram\newline
\begin{enumerate}[label=(\alph*)]
    \item \textbf{\(E<<2mgl\)}
\end{enumerate}
\begin{eqnarray}
\label{eq3}
1&=&\frac{2mgl\sin^2\frac{\theta}{2}}{E}+\frac{L^2}{2ml^2E}\\
\label{eq3.1}
1 &=&\frac{mgl\theta^2}{2E}+\frac{L^2}{2ml^2E}
\end{eqnarray}
It is possible to rewrite equation \hyperref[eq3.1]{2} into \hyperref[eq3]{3} using the small angle approximation - \(\sin\theta\approx\theta\). This is possible noting that when \(E=\frac{1}{2}m \omega^2 l^2 \theta^2\) (energy of harmonic oscillator) is small, \(\theta\) also becomes small. Physically, this condition represents a small oscillation around its stable point.\newline
\begin{enumerate}[label=(\alph*)]
\setcounter{enumi}{1}
    \item \textbf{\(E=2mgl\)}
\end{enumerate}
 \begin{equation}
    \label{eq 4}
         L=m\sqrt{2gl^3}\cos\frac{\theta}{2}
    \end{equation}
Equation \hyperref[eq4]{4} may be obtained from equation \hyperref[energy 1]{1}, where we substitute the relation \(E=2mgl\). This marks a transition from the first phase to the second phase\footnote{Refer to the final sentence in \hyperref[Abstract]{Abstract}}\newline
 \begin{enumerate}[label=(\alph*)]
    \setcounter{enumi}{2}
    \item \(E>2mgl\)
    \end{enumerate}
\begin{equation}
    \label{eq 5}
    L^2=2ml^2[E-2mgl\sin^2\frac{\theta}{2}]
    \end{equation}
Equation \hyperref[eq 5]{5} may be obtained by simply rewriting equation \hyperref[energy 1]{1} without making any approximations. This equation can be used to draw the phase diagram of phase 2 in the pendulum system. Physically, this condition is reached when the pendulum can perform a full rotation\footnote{All phase diagrams can be seen in section \hyperref[2.7]{2.7}}

\subsection{Slack Analysis}
\label{subsection2.2}
\subsubsection{Conditions}
For this part only, we consider a string pendulum. This time, specific boundary conditions need to be fulfilled to perform a complete circular motion. \newline
We consider the following constraint
\[ml\dot{\theta}^2+mg\cos\theta>0\]
Insertion into the energy equation of the pendulum yields
\[E>mgl-\frac{3}{2}mgl\cos\theta\]
Notice how if \(E=\frac{5}{2}mgl\), slack will never occur. Rewriting in terms of E, yields
\begin{equation}
\label{eq 6}
\cos\theta>\frac{2}{3}[1-\frac{E}{mgl}]
\end{equation}
This means before the lower bound is reached, the strings will remain taut. 
\newline
We may obtain the angular speed of the pendulum at this instant.
\begin{equation}
\label{eq 7}
\dot{\theta}^2=\frac{2g}{3l}[-1+\frac{E}{mgl}]
\end{equation}
\subsubsection{Slack Time}
We find the time required by the string to become taut again. Note that the pendulum will undergo parabolic motion during this time range, and the following equation must be fulfilled for it to become taut again\footnote{Due to the complexity in the threaded lato-lato's motion, we will refrain from using it in our analysis. Instead, the focus will lie on a stick-based lato-lato.}. 
\begin{equation}
\label{eq 8}
y'^2+x'^2=l^2
\end{equation}
Where \(r'=(x',y')\) is the instantaneous position of the object. We define \(y'=-l\cos\theta+v_0 t\sin\theta -\frac{1}{2}gt^2\) and \(x'=l\sin\theta+v_0 t\cos\theta \). 
Insertion allows us to get the nontrivial equation
\[\frac{1}{4}gt^2-v_0 gt\sin\theta +(v_0^2+gl\cos\theta)=0\]
\[gt=2v_0 \sin\theta \pm \sqrt{-v_0^2 \cos^2\theta-g^3l\cos\theta}\]
inserting the value of \(\cos\theta=\frac{-v_0^2}{gl}\), will allow us to write 
\begin{equation}
\label{eq 9}
t=\frac{4v_0 \sin\theta}{g}
\end{equation}
From this point onward, the velocity component in the direction of the string will be eliminated, leaving the tangential component. This process will result in energy loss.
\subsection{Pendulum Dynamics}\label{subsectionpd}
In this part, we will be deriving the most important formula that is used in the entirety of the paper. This is the equation of motion of the pendulum, with its free end driven by an oscillating force (\(\frac{F}{m}=-a_0 \omega^2 \cos\omega t\))\footnote{\(a_0\) shows the amplitude of the motion undergone by the pendulum's free end and \(\omega\) shows the angular frequency of the oscillatory motion}. We consider the frame of the oscillating free end, such that the motion of the spheres is exactly circular.
\begin{eqnarray}
    E&=&mgl(1-\cos\theta)+ml^2\frac{\dot{\theta}^2}{2}+\int-ma_0 \omega^2 \cos\omega t dy \label{eq 2.1} \\ 
   \frac{dE}{dt}&=&0 \label{eq 2.2} \nonumber \\  
   0&=&mgl \sin\theta \dot{\theta} +ml^2\ddot{\theta}\dot{\theta}-ma_0l \omega^2 \cos\omega t \sin\theta\dot{\theta} \label{eq 2.3} \nonumber \\
-\ddot{\theta}&=& (g-a_0 \omega^2 \cos\omega t)\sin\theta \label{eq2.4}
\end{eqnarray}
Equation \ref{eq2.4} is known as the \textit{Mathieu’s equations}\cite{bib2-pendulum}, having a general solution of\cite{bartucelli2}:
\begin{equation}
   u(t)= e^{i\mu_0 t}p_0(t)
\end{equation}
In general, \(\mu_0\in\mathbb{C}\). If \(\mu_0\in\mathbb{R}\), the solution is particularly bounded. 
\subsection{Kapitza Model}\label{subsection 2.4}
To perform code proof-testing later on in \textbf{section 4}, we will consider several constraints to make equation (10) analytically solvable, which are\cite{butikov2021kapitza}:
\begin{enumerate}
\item Small value of \(a_0\)
\item Fast oscillation frequency \(\omega\)
\item This way we may rewrite \(\theta=\gamma+\beta\), where \(\gamma\) is the slow varying term, with large amplitude and \(\beta\) is the opposite of \(\gamma\)
\end{enumerate}

We first try to obtain the value of \(\beta\). We note that the second derivative of \(\gamma\) is way smaller than that of \(\beta\). This way, we may expand the terms, hence ending up with
\begin{equation}
 (g-a_0 \omega^2 \cos\omega t)\sin\gamma=\ddot{\beta}l
\end{equation}

noting that \(\omega^2a_0>>g\), we end up with
\begin{equation}
    \beta=\frac{a_0}{l}\sin\gamma\cos\omega t
\end{equation}

Next, we iterate the obtained \(\beta\) on the equation of motion to get \(\gamma\). We will neglect terms of the order \(y^3\) and so on
\begin{equation}
    g\sin\gamma+g\cos\gamma \beta -a_0\beta \omega^2 \cos\omega t\cos\gamma=-\ddot{\gamma}
\end{equation}

Noting the 2nd approximation condition, we may average the previous function to obtain:
\begin{equation}
    \frac{g\sin\gamma}{l}+\frac{a_0^2\omega^2}{l^2}\frac{\sin2\gamma}{4}=-\ddot{\gamma}
\end{equation}

Next, we will find the average moment of forces acting on the pendulum
\[\tau=m\ddot{\gamma}l^2\]
\begin{equation}
\tau=-(mgl\sin\gamma +ma_0^2\omega^2\frac{\sin2\gamma}{4})
\end{equation}

Then, we define a scalar potential due to this torque
\begin{equation}
  V(\gamma)=-\int\tau d\gamma
\end{equation}
\begin{equation}
 V(\gamma)=mgl(1-\cos\gamma)+ma_0^2\omega^2\frac{\sin^2\gamma}{4}
\end{equation}
We continue by analyzing several stability options.
\[\frac{dV(\gamma)}{d\gamma}=0\]
from there, we obtain \(\gamma_1=0\), \(\gamma_2=\pi\), \(\cos\gamma_3=-\frac{2gl}{y_0^2\omega^2}\)\footnote{We define \(V_{max}\) as the value of the scalar potential \(V\) at which \(\gamma=\gamma_3\).}. The stability for \(\gamma=\pi\) only works if \(\gamma_3\) has a real solution, such that \(\frac{d^2V}{d\gamma^2}\) at that point \(<0\).
\newline

Here, we can define the energy of the system in the moving frame as 
\begin{equation}
    E'=V+\frac{ml^2\dot{\theta^2}}{2}
\end{equation}

defining \(L\),
\begin{equation}
L^2=2ml^2(E'-mgl(1-\cos\gamma)+ma_0^2\omega^2\frac{\sin^2\gamma}{4})
\end{equation}

\subsection{Energy Loss}
We consider an energy loss proportional to \(e^2\), where \(e<1\) represents the coefficient of restitution of the two bobs. Due to this, we can write \(E_n\), that is the energy to the nth collision as \(E_n=e^nE_0\).
\[L^2=2ml^2[E_n-2mgl\sin^2\frac{\theta}{2}]\]
Notice that the system will lose kinetic energy after several collisions, which means that additional energy must be given every time energy is dissipated.
For every energy loss, the following \(\delta W\) must be given.
\begin{equation}
\delta W=\Delta E_n=E_{n-1}[e-1]
\end{equation}
We can now see the importance of giving additional work to keep the pendulum at its original energy state. We may do this by lifting the pendulum system up and down, with the power defined as:
\begin{equation}
<P>=<\vec{F}\cdot\vec{v}>
\end{equation}

\subsection{Tornado Play Style}\label{subsec2.6}
Apart from the regular pendulum play style, the \textit{lato-lato} can also be played less conventionally. To model this style, we refer to \hyperref[fig:enter-label1]{Figure 1}. In this model, we will consider the human finger as a cylindrical wheel having radius \(R\), where (\(r<<R<<l_0\)), rotating at a constant angular velocity \(\omega\). 

To analyze the dynamics of the system we shall first consider the movement of the pendulum in the rotating frame \(\omega\), and then we will transform the kinematic properties back into the inertial lab frame. This will ease the maths involved. The following formula for transformation will be used\cite{greenwood}
\begin{equation}
    \vec{v}_{lab}=\vec{v_0}+\vec{\omega} \times \vec{r}+\vec{\dot{\rho}}_r
\end{equation}
\begin{equation}\label{cool}
\vec{a}_{lab}=\vec{a_0}+\vec{\omega}\times(\vec{\omega}\times \vec{r})+\dot{\vec{\omega}}\times \vec{r}+ 2\vec{\omega} \times \vec{\dot{\rho}}_r+\vec{\ddot{\rho}}_r
\end{equation}
where \(\vec{v_0}\) defines the velocity of the origin in the lab frame, \(\vec{\omega}\) defines the angular velocity of the rotating frame, \(\vec{\dot{\rho}}_r\) defines the velocity of the object in the rotating frame, and lastly \(\vec{\ddot{\rho}}_r\) defines the acceleration experienced by the object in the rotating frame. 
\subsubsection{Wrapping}
For the initial part, we will derive the time taken for the pendulum to be fully retracted until \(l=0\). This can be done by giving the sphere an initial momentum such that the string winds around our finger. By assuming the initial speed given is such that \(v_0^2>>gl\), we may ignore the effects of gravity. Therefore, one could write
\begin{equation}
v_0=\dot{\theta}l=constant
\end{equation}
where \(\theta\) is the wrapping angle and \(l=l_0 - R\theta\). Solving the above differential equation yields the following analytical result
\begin{equation}
    t=\frac{l_0\theta-R\frac{\theta^2}{2}}{v_0}
\end{equation}
inserting \(\theta=\frac{l_0}{r}\) (under the approximation \(r<<l\)), will yield the final result \(t_R=\frac{l_0^2}{2v_0R}\).
\subsubsection{Unwinding}
This part will now use the formulas provided at the beginning of this subsection. The model used here is that we quickly rotate our finger with a constant angular velocity \(\omega\). We again assume that gravity is negligible. We first notice that considering a frame rotating at angular speed \(\Omega=\omega+\dot{\theta}\) will be much easier. This is because, by considering this frame, we will be given a system where the pendulum's string just changes in length, without any rotational motion. In this frame, the kinematic properties are simply :
\begin{equation}
    \vec{\dot{\rho}}_r = \dot{l} \hat{l}=R\dot{\theta} \hat{l}
\end{equation}
\begin{equation}
    \vec{\ddot{\rho}}_r = \ddot{l} \hat{l}=R\dot{\theta} \hat{l}
\end{equation}
Subsequently, insertion into equation \hyperref[cool]{Equation 23}
\begin{eqnarray}
    \vec{a}_{lab}&=&-\Omega^2\vec{R}-\Omega^2\vec{l}+\ddot{l}\frac{l}{R} \hat{e_t}+ 2\omega\dot{l}\hat{e_t}+\ddot{l}\hat{l} \nonumber\\
    (\vec{a}_{lab})_{\hat{e_t}}&=&-(\omega+\frac{\dot{l}}{R})^2{R}+\ddot{l}\frac{l}{R}+2\Omega\dot{l}\nonumber\\
    (\vec{a}_{lab})_{\hat{e_t}}&=&\frac{\dot{l^2}}{R}+\frac{\ddot{l}l}{R}-\omega^2R
\end{eqnarray}
The vector \(\hat{e_t}=\frac{\vec{e_t}}{e_t}\), represents the direction perpendicular to the string \(\hat{l}\). Notice that the force in this direction is negligible (gravity), therefore we may immediately set it to 0. The equation can be turned into a perfect integral
\begin{equation}
    0=\frac{d}{dt}(l\dot{l})-\omega^2R^2
\end{equation}
integrating both sides by \(dt\), and setting the initial condition \(l_0=0\), allows us to write
\begin{eqnarray}
    l\dot{l}&=&\omega^2R^2t\nonumber\\
    \int_{0}^{l} l\,dl &=& \frac{1}{2}\omega^2R^2t^2\nonumber\\
    l&=&\omega Rt
\end{eqnarray}
It turns out, the \(l(t)\) function is linear, hence we may write unwinding time \(t_u\) as
\begin{equation}
    t_u=\frac{l_0}{\omega R}
\end{equation}
\subsubsection{Total time}
We assume motion starts from unwinding, and when \(l\) reaches \(l_0\), the \(\omega\) value is immediately set to 0. Hence, by combining the total time \(T=t_R+t_u+t_t\) (third term = transition time), we can approximate the period of each "tornado" motion as 
\begin{equation}
    T=\frac{l_0}{R}(1+\frac{\omega l_0}{2v_0})+\frac{\pi}{v_{\hat{e_t}}}
\end{equation}
Noting that \(\vec{v}=-(\omega{R})\hat{l}+\omega l\hat{e_t}\), we can find the velocity of the sphere when the length has reached \(l_0\), \(v_{\hat{e_t}}=\omega l_0\). The radial velocity can be ignored because when the string quickly goes back to being taut - noting (\(R<<l\)) - the radial component just vanishes. \(v_0\) defined previously is also equal to \(v_{\hat{e_t}}\), due to the periodicity defined. Our expression simplifies into:
\begin{equation}
    T=\frac{3l_0}{2R}+\frac{\pi}{\omega}
\end{equation}
\label{2.7}\subsection{Preliminary Figures}
 
\begin{figure}[H]
    \centering
    \includegraphics[width=0.7\textwidth,clip=]{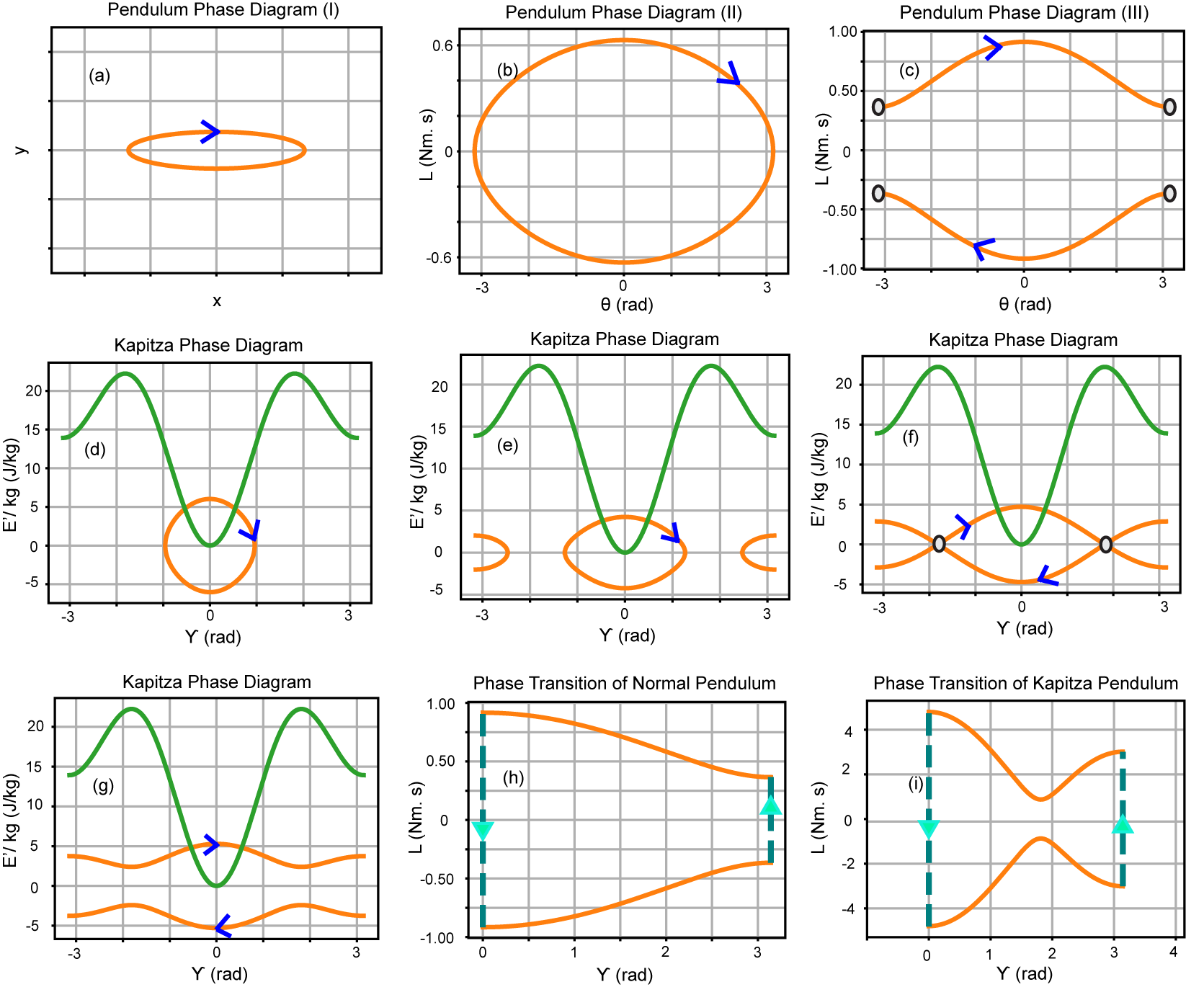}
    \caption{The orange contours in the plots show the phase of the pendulum system, whereas the green contour shows the potential barrier experienced by the pendulum in the kapitza model. Figures \textbf{(a)},\textbf{(b)}, and \textbf{(c)} show the phase diagram of the special cases, where its equations are given in \hyperref[subsection1]{subsection 1}, representing a simple mathematical pendulum. Meanwhile, figures \textbf{(d)}, \textbf{(e)},\textbf{(f)}, and \textbf{(g)}, represent the phase diagram for the kapitza model, that analyzes the conditions \(0<E'<2mgl\), \(2mgl<E'<V_{max}\), \(E'=V_{max}\), and \(E'>V_{max}\) respectively. The orange contour in the four figures represents a plot of the system's potential barrier, whereas the red contour shows the phase diagram \(E'\) vs \(\theta\) of the kapitza model. Finally, \textbf{(h)} and \textbf{(i)} show the phase transitions that occur for a simple pendulum and the kapitza model. Both are done at the most extreme case of each respective condition.  }
    \label{Figure 2}
\end{figure}
In this part, we have provided the plots of trivial single pendulum phase diagrams, as well as other figures that might aid in illustrating \hyperref[sec2]{Section 2}.

\section{Application in Lato-lato}\label{sec3}
Having derived all the necessary equations we may now solve the equation of motion (\hyperref[eq2.4]{equation 10}) numerically. Notice how this equation shows how the lato-lato is usually moved around by the player. Since equation 10 only shows the motion of a single pendulum, it is required that we add specific constraints, characterizing the geometry of a double pendulum system. The constraints are as follows:
We apply 2 boundary conditions at \(\theta=0\) and \(\theta=\pi\). Forcing, \(\dot\theta_{(0)}\) and \(\dot\theta_{(\pi)}\) to always be reversed. 
\begin{equation}
    (\dot\theta_{0})_i=-(\dot\theta_0)_{i+1}
\end{equation}
\begin{equation}
    (\dot\theta_{\pi})_i=-(\dot\theta_{\pi})_{i+1}
\end{equation}
This represents the almost elastic collisions between the plastic spheres, which immediately models the momentum conservation law between the plastic spheres.
The codes needed to solve the equation numerically and to obtain plots of the system are all provided by chatGPT. These codes can be seen in the \hyperref[secA1]{Appendices}.

\section{Code Proof-testing in Kapitza Model}\label{sec4}
To test the performance of the code, we simulate a special case called the Kapitza model, as in \hyperref[subsection 2.4]{section 2.4} To do this, we just need to use code (b) (\(\theta\) as a function of time) and insert some numeric constants which are in agreement with the approximations in \hyperref[subsection 2.4]{section 2.4}. It is also important to note that the equation of motion - equation \hyperref[eq2.4]{11} - of the pendulum is strictly independent of the bob's mass, hence for convenience, we may as well set \(m= 0.020 \)kg. In addition, to cover all ranges of mass, we define the specific energy \(E'\) as the energy per unit mass (kg).
We use the following numeric values:
\begin{multicols}{2}
    \begin{enumerate}[label=(\alph*)]
	\item \(a_0=0.01 \) m
	\item \(l=0.25 \) m
	\item \(g=9.81\) \(m/s^2 \)
        \item \(\omega=100\) rad/s
	\item \(\theta_{(0)}=1.5 \) rad
	\item \(\dot\theta_{(0)} =0 \) rad/s
 \end{enumerate}
\end{multicols}
\begin{figure}[H]
    \centering
    \includegraphics[width=0.6\textwidth,clip=]{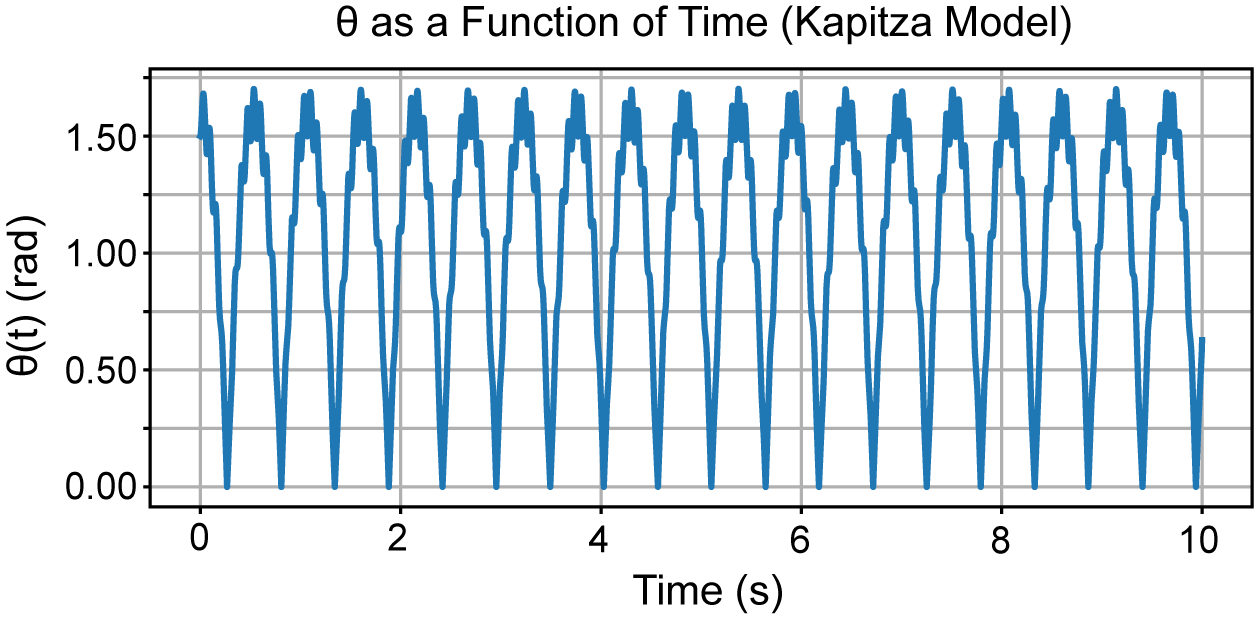}
    \caption{We notice the \(\theta\) may be decomposed just as our approximation in section 2.4 (\(\theta\approx\gamma+\beta\))}
    \label{fig:enter-label}
\end{figure}
From here, it is safe to say that the codes provided should work just fine
\section{Results in Lato-lato}\label{sec5}
Now that we have the codes that represent the \textit{lato-lato's} mechanics, we can analyze the reason why playing the lato-lato is difficult. In general, we have observed that there are 2 main phases in the system's motion. To analyze we will use the 4 codes we have attached in \href{https://github.com/FansenCandra/All-Lato-Lato-Code}{GitHub}\cite{GitHub}. 
\subsection{Maximum Amplitude}

The code provided in \cite{GitHub} will plot a graph of the maximum amplitude vs \(\omega\). From this graph, we can see that there are 2 phases. The first phase represents the normal pendulum motion and the second phase represents the condition at which the spheres collide at both \(\theta=0\) and \(\theta= \pi\) rads.
\begin{figure}[H]
    \centering
    \includegraphics[width=0.88\textwidth,clip=]{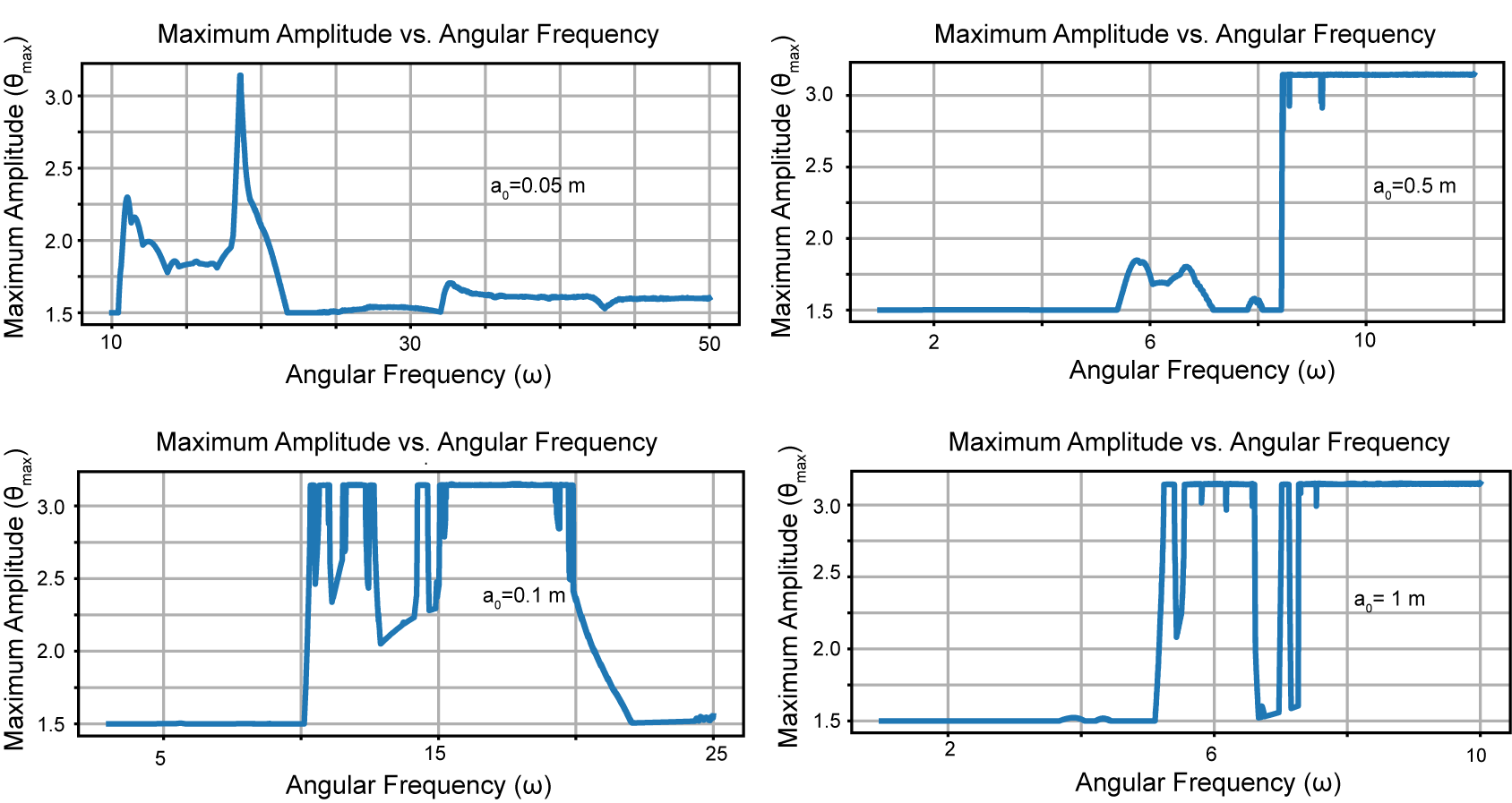}
    \caption{The maximum amplitude is only considered within the 3-second range after the motion starts. This is done to imitate the way people usually play the lato-lato. The figures are plotted for \(a_0= 0.05\) m, \(a_0=0.1\) m, \(a_0=0.5\) m and \(a_0=1\) m.}
    \label{fig:enter-label}
\end{figure}

The three graphs represent the plots for different \(a_0\). The graph in the upper left corner shows the plot for \(a_0=0.5\) m, the one in the upper right corner is the zoomed version of the latter, the one in the lower left shows for \(a_0=0.1\) m, and in the lower right corner shows for \(a_0=1\) m.
Notice how different \(a_0\) yield different maximum \(\omega\) graphs. \newline Generally, one may observe that when the driven amplitude \(a_0\) is increased, the minimum driven angular frequency \(\omega\) needed to reach phase 2 is lower. For lower amplitudes on the other hand, specific \(\omega\) should be maintained to reach phase 2. This strengthens the argument as to why the lato-lato is a difficult game. 

\subsection{\(\theta\) as a function of time}
In this part, we can see the graph that represents the polar coordinates of the lato-lato as a function of time. Through these plots, we should be able to get a general feel of the system's motion. 
\begin{figure}[H]
\centerline{\hspace*{0.015\textwidth}
         \includegraphics[width=0.5\textwidth,clip=]{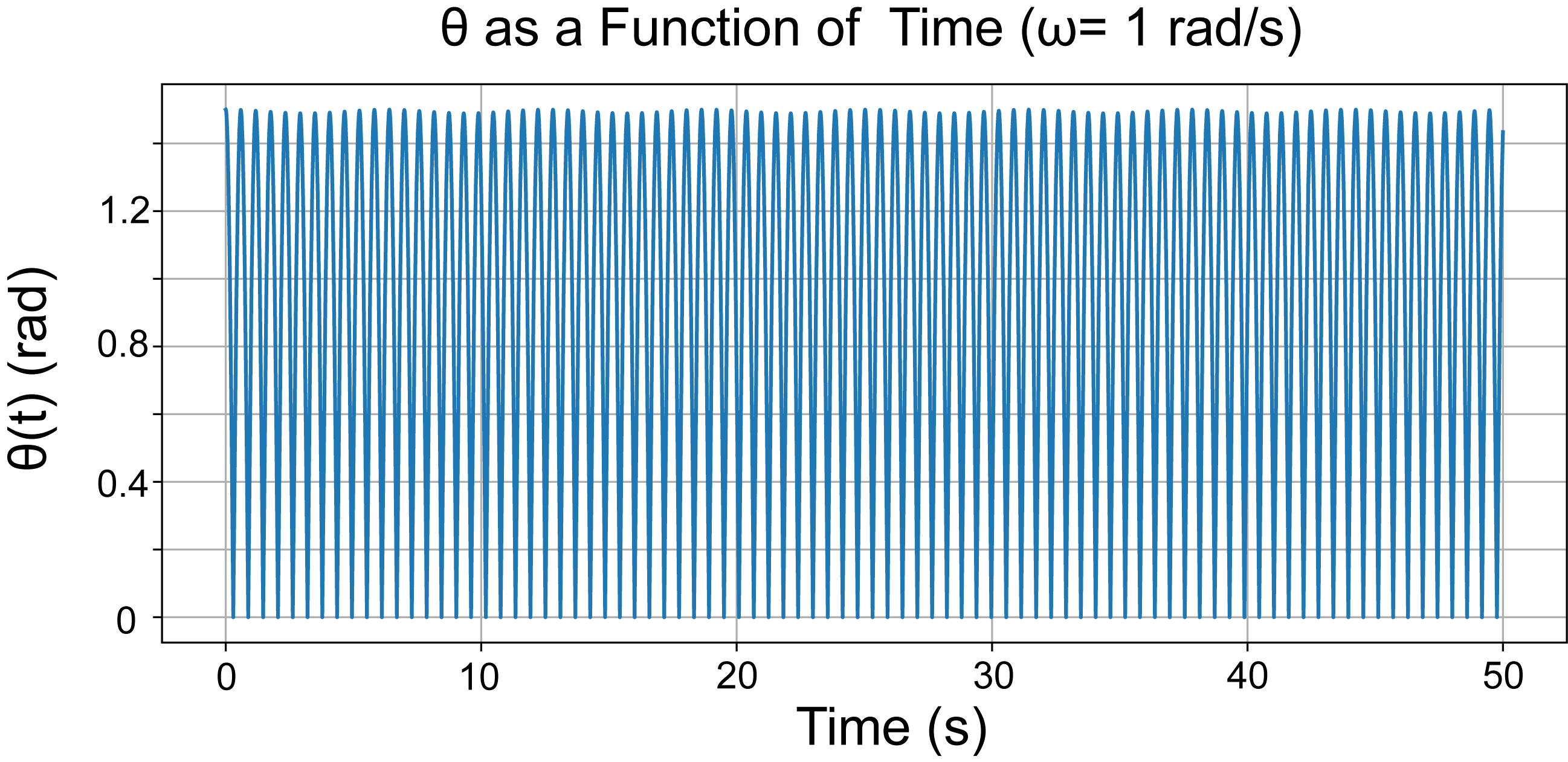}
         \hspace*{0.005\textwidth}
         \includegraphics[width=0.5\textwidth,clip=]{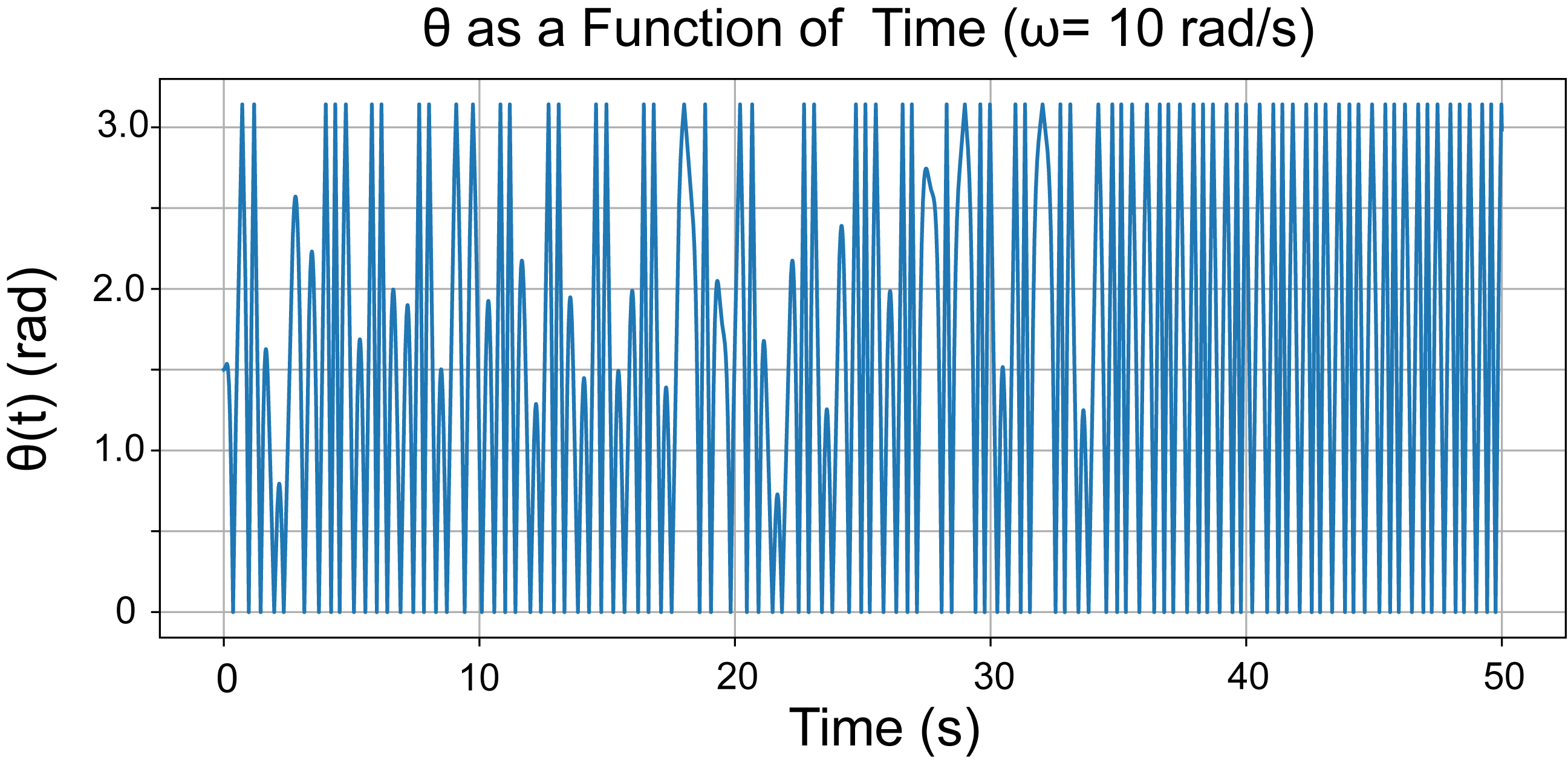}
        }
\vspace{-0.35\textwidth}   
\centerline{\Large \bf     
\hspace{0.0 \textwidth}  \color{white}{(a)}
\hspace{0.415\textwidth}  \color{white}{(b)}
   \hfill}
\vspace{0.31\textwidth}    
\caption{Figure on the left shows \(\theta\) as a time function for \(\omega= 1\) rad/s and the one on the right shows for \(\omega=10\) rad/s. For the smaller \(\omega\), we have a sinusoidal amplitude function and a less periodic wave function for \(\omega>10\) rad/s} 
\label{F-4panels}
\end{figure}
Figure 4 shows the \(\theta\) vs time function for \(a_0=0.5\) m. In general, we can see that increasing the \(\omega\) whilst keeping the \(a_0\) constant will help the \textit{lato-lato} reach the 2nd phase. However, this will not be the case if \(a_0\) is too small.\newline 
At lower angular frequencies, we may observe an amplitude undergoing sinusoidal change. Furthermore, at higher angular frequencies, the second phase is easily reached, though the motion does not seem as periodic as the latter. 
\subsection{Energy as a function of time and \(\theta\)}
\begin{figure}[H]
    \centering
    \includegraphics[width=0.92\textwidth,clip=]{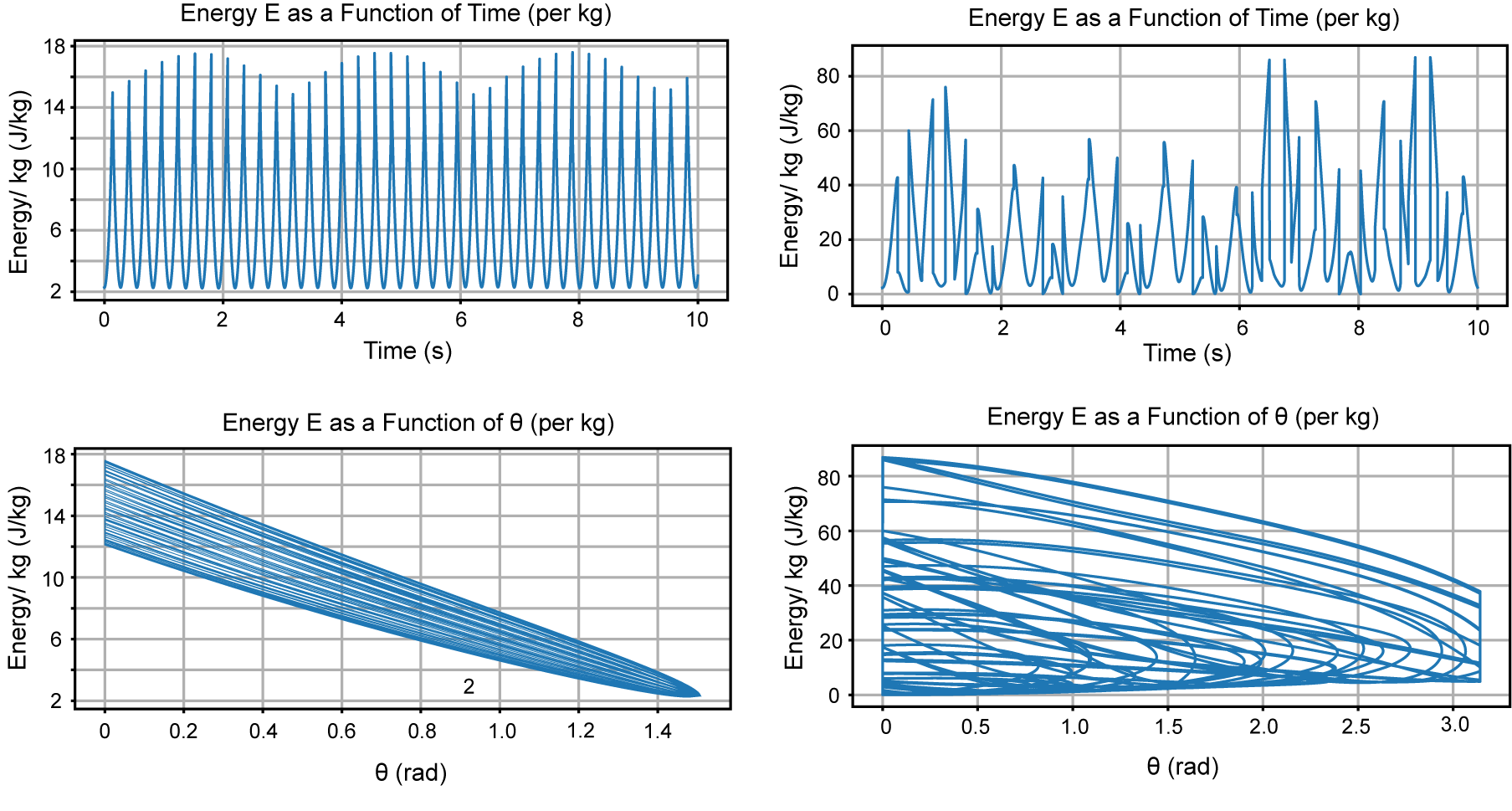}
    \caption{Figures on the left show one of the pendulum's energy as a time function and \(\theta\) for \(\omega= 1\) rad/s and the on the right shows for \(\omega=10\) rad/s. For the smaller \(\omega\), we have a periodic graph, whereas the larger one shows a rather irregular progression. All the graphs are plotted at \(a_0=0.5\) m}
    \label{Figure 6}
\end{figure}
Through the plots in this section, we should be able to understand the energy transfer\footnote{Energy transfer here, refers to how much work is being transferred from the person's hand to the lato-lato system.} that occurs between the system and the driving force. We will yield plots that show how the energy transfer varies as a function of the driven \(\omega\). Consequently, one may compare how the energy of the system relates to the lato-lato's position. 
\newline\newline
First, it is important to notice that at lower angular frequency \(\omega\), the energy indeed has an amplitude varying periodically, similar to the \(\theta\) plot we have obtained. Likewise for higher \(\omega\), a fluctuating pattern can be seen. Moreover, notice how the energy transfer is larger for a higher \(\omega\). The energy reached at \(\theta=0\)  may yield relatively high values, reaching a maximum of  \(E_{max}=17.60\) J/kg and minimum \(E_{min}=2.23 \) J/kg for \(\omega= 1\) rad/s. A less orderly graph can be seen, with a maximum energy of \(E_{max}=86.88 \) J/kg and \(E_{min}=0.01 \) J/kg for the higher \(\omega=10\) rad/s. This explains why it is easier for the system to reach the 2nd phase when given a higher angular frequency, recalling that the energy transfer is higher in the high angular frequency case. 

\subsection{System's stability}
The mesh plots provided in this section will directly show us the phase conditions in the lato-lato's phase diagram, that is whether it reaches the 2nd phase or not. The mesh plots will give a diagram representing the amplitude conditions vs the initial boundary conditions \(\theta_0\) and \(\dot\theta_0\). In this section, we will be able to analyze which sort of amplitudes will give the easiest condition required to reach the 2nd phase of the motion.
We will compare three driven amplitudes, which are \textbf{(1)} \(a_0=0.1 m\), \textbf{(2)} \(a_0=0.5 m\), and \textbf{(3)} \(a_0=1 m\). 
\subsubsection{\(a_0=0.05 m\)}\label{sec5.4.1}
We can see the yellow dot reaches the 2nd phase at around \(\omega= 13-16\) rad/s\footnote{People usually play the lato-lato at this angular frequency, equivalent to a frequency of about 2-3 Hz)}. 
From the mesh plots, we can also see more unstable\footnote{"unstable" refers to not being able to reach phase 2 (colored red).} regions (there is a larger blue area in between the red ones), especially at around 15-18 rad/s. Due to the irregularity in the region distribution, we may infer that relatively small changes in the angular frequency can cause huge changes in the system's mesh plots. For the purpose of our analysis, we will focus on the transition experienced by the yellow dot at around \(\omega=16.25\) rad/s, and \(\omega=13\) rad/s. \newline
Taking the average of the lower (13 rad/s) and upper limit (16.25 rad/s) of the typical \(\omega\) values, yields \(\omega=14.625 \) rad/s. This gives a \(\delta_{low}\)\footnote{Defined as \(\frac{\Delta\omega}{\omega}\). \(\Delta\omega\) is calculated as \(\omega_f -\omega_0\), where \(\omega_f\) represents the \(\omega\) at which the point experiences a transition in phase. A lower value of \(\delta\) shows a larger sensitivity} of 12.5\(\%\) and \(\delta_{high}\) of about  11.1\(\%\). At this \(\omega\), we get a relatively stable condition. Therefore, to reach phase 2 in the lato-lato's motion, one has to maintain a specific angular frequency and initial conditions. 

Here, we have made the assumption that the changes are done at the yellow dot. This means if the error is done at random initial conditions \(\theta_0\) and \(\dot{\theta}_0\), for instance at (\(1\) rad, \(-1.5\) rad/s ), the system will no longer be able to reach the desired phase 2, in 3 seconds.
\begin{figure}[H]
    \centering
    \includegraphics[width=0.825\textwidth,clip=]{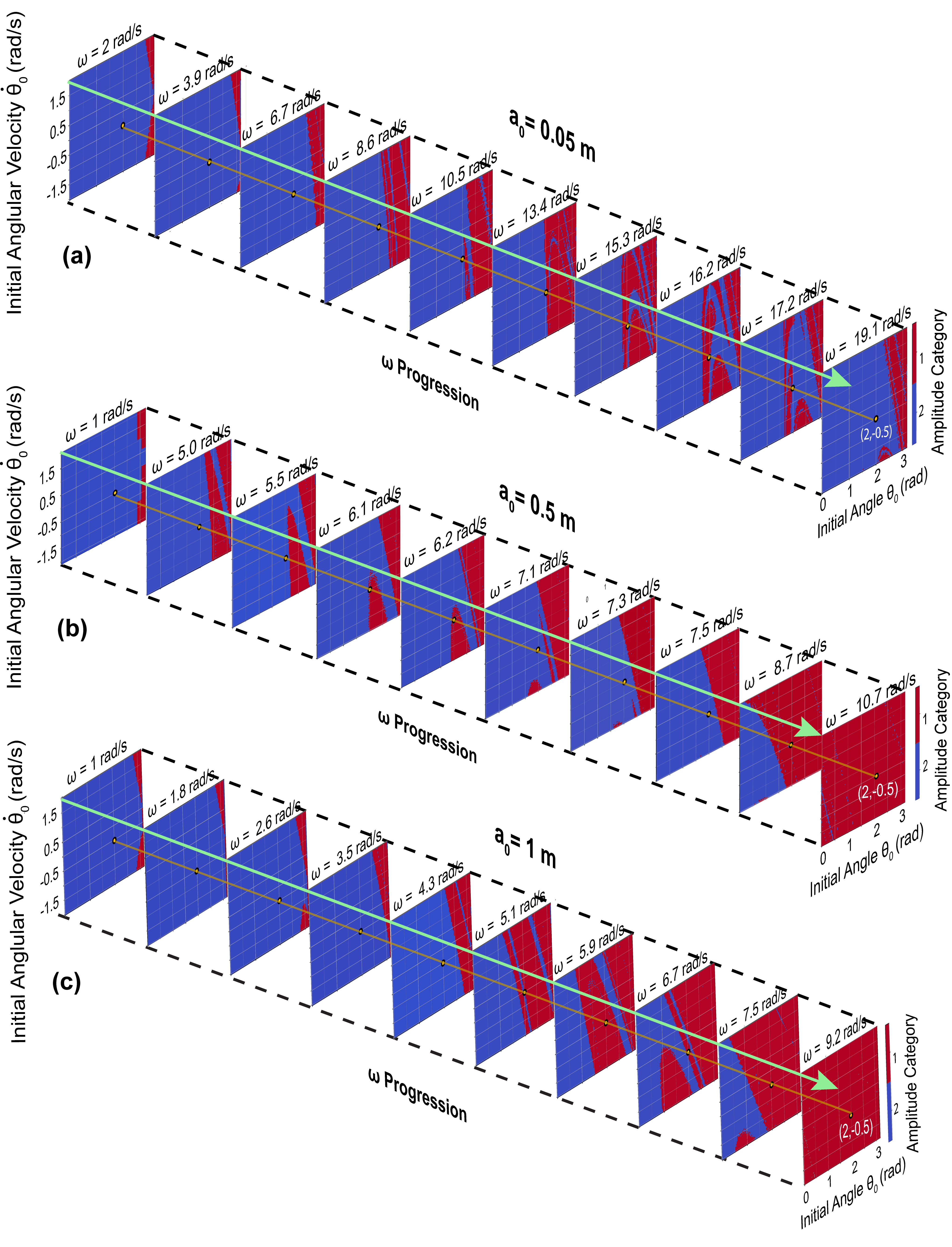}
    \caption{Mesh plot progression for \(a_0= 0.05 m\), \(a_0= 0.5 m\) and \(a_0= 1 m\) showing \textit{Amplitude Conditions} for different \(\omega\). Phase 1 in the mesh plots is denoted by the color blue and phase 2 by the color red. The yellow dot shows the observed condition, that is at \(\theta_0=2\) rad and \(\dot{\theta_{0}}=-0.5\) rad/s}
    \label{fig:enter-label}
\end{figure}
\subsubsection{\(a_0=0.5 m\)}

 Notice at a very high \(\omega\), the system will reach the 2nd phase regardless of the boundary conditions. 
Focusing on the yellow dot, it can be seen the system starts to enter the 2nd phase at \(\omega=6.1 \)rad/s and \(\omega=6.5  \) rad/s. Subsequently, it enters the 2nd phase again at \(\omega=8 \)rad/s, from which it stays there for any \(\omega\) larger than the former. If we consider the \(\omega=6.3 \) rad/s, the \(\delta\) towards \(\omega=6.1 \) rad/s is about \(3.8 \%\) and \(3.17 \%\) towards \(\omega=6.5\) rad/s. This shows that at the hill-like contour, the system is not very stable when compared with the case \((a_0,\omega)=(0.05 \) m, \(14.63   \) rad/s). However, specific initial conditions, such as (3,1.5), allow a very stable condition, as shown in the mesh plot progression.\newline
\subsubsection{\(a_0=1 m\)}

Even though the numeric value for this consideration may be considered unrealistic, we may still get a decent analysis by considering this case. Notice how the small budge in \textbf{Figure 3} now reaches the maximum \(\theta=3.14\) rad. This means that when we increase the driven amplitude \(a_0\), it becomes easier to reach the 2nd phase of the \textit{lato-lato's} motion. The yellow dot quickly reaches the 2nd phase at \(\omega=5\) rad/s, and for all \(\omega\) bigger than 5, it stays in the 2nd phase. Moreover, it becomes clearer that the considered initial condition reaches the “all-red” part of the diagram more quickly, as expected when the amplitude is increased. 

Therefore, it is better to just play at a high frequency, because that way we can guarantee a larger energy transfer\footnote{Energy transfer here, refers to the energy transfer from the player's hand to the \textit{lato-lato} system} (stable at phase 2).

\section{Conclusion}\label{sec13}
Having analyzed the conditions for reaching phase 2 of the \textit{lato-lato's} motion, we may finally infer that there are certain driven angular frequencies and amplitudes required to maintain its motion. These characteristics may be sensitive as observed in \hyperref[sec5.4.1]{Section 5.4.1}, for \(a_0= 0.05 m\). Even at a very specific \(\omega=14.63 \) rad/s, we only get a \(\delta\) of as much as 12\%, which can be considered extremely sensitive. As a result, subtle changes at the incorrect moment may prevent continuous collisions between the bobs. Through the analysis of phase diagrams, we have qualitatively proven that playing the \textit{lato-lato} is indeed difficult. Furthermore, throughout this paper, we have only considered the dynamics of a stick-based pendulum, which is easier to control than the one made of string. Therefore, trained muscle memory and experience are some of the necessities required if we wish to master this game.

\backmatter

\bmhead{Acknowledgments}
We would like to deliver our deepest appreciation to Sandy Adhitia Ekahana for the spark of ideas\cite{sandyadhitia}. Further, this paper would not have been made possible without the codes required for the numerical analysis which are provided by the AI ChatGPT.




\newpage

\bibliography{sn-bibliography}

\end{document}